\documentclass[
 reprint,
superscriptaddress,
nofootinbib,
 amsmath,amssymb,
 aps,
 prl,
floatfix,
]{revtex4-1}

\usepackage{graphicx,color}
\usepackage{dcolumn}
\usepackage{bm}
\usepackage{hyperref}
\usepackage{physics}
\usepackage{soul}
\usepackage{cancel}
\usepackage{mathtools}
\usepackage{cprotect}
\usepackage{enumitem}
\usepackage[normalem]{ulem}
\usepackage{comment}
\newcommand{\etal}{{\it et al.}}

\usepackage[multidot]{grffile}

\usepackage[caption=false]{subfig}

\usepackage[dvipsnames]{xcolor}
\definecolor{C0}{RGB}{31, 119, 180}
\definecolor{C1}{RGB}{255, 127, 14}
\definecolor{C2}{RGB}{44, 160, 44}
\definecolor{C3}{RGB}{214, 39, 40}
\definecolor{C4}{RGB}{148, 103, 189}
\definecolor{C5}{RGB}{140, 86, 75}

\begin{document}

\title{The Gravitational Wave Memory from Binary Neutron Star Mergers} 

\author{Jamie Bamber}
\affiliation{%
Department of Physics, University of Illinois Urbana-Champaign, Urbana, IL 61801, USA
}
\author{Antonios Tsokaros}
\affiliation{%
Department of Physics, University of Illinois Urbana-Champaign, Urbana, IL 61801, USA
}%
\affiliation{National Center for Supercomputing Applications, University of Illinois Urbana-Champaign, Urbana, IL 61801, USA}
\affiliation{Research Center for Astronomy and Applied Mathematics, Academy of Athens, Athens 11527, Greece}
\author{Milton Ruiz}%
\affiliation{Departament d’Astronomia i Astrof\'{i}sica, Universitat de Val\`{e}ncia, C/ Dr Moliner 50, 46100, Burjassot (Val\`{e}ncia), Spain}
\author{Stuart L. Shapiro}
\affiliation{%
Department of Physics, University of Illinois Urbana-Champaign, Urbana, IL 61801, USA
}%
\affiliation{
Department of Astronomy, University of Illinois Urbana-Champaign, Urbana, IL 61801, USA
}
\affiliation{National Center for Supercomputing Applications, University of Illinois Urbana-Champaign, Urbana, IL 61801, USA}
\author{Marc Favata}
\affiliation{Department of Physics \& Astronomy, Montclair State University, 1 Normal Avenue, Montclair, NJ 07043, USA}
\author{Matthew Karlson}
 \affiliation{Department of Physics \& Astronomy, Montclair State University, 1 Normal Avenue, Montclair, NJ 07043, USA}
 \affiliation{Department of Mechanical Engineering \& Materials Science, University of Pittsburgh, Pittsburgh, PA 15260}
\author{Fabrizio Venturi Pi\~{n}as}
\affiliation{Departament d’Astronomia i Astrof\'{i}sica, Universitat de Val\`{e}ncia, C/ Dr Moliner 50, 46100, Burjassot (Val\`{e}ncia), Spain}

\date{\today}

\begin{abstract} 
The gravitational wave  signal produced by the merger of two compact objects includes both an oscillatory transient and a non-oscillatory part, the so-called memory effect. This produces a permanent displacement of test masses and has not yet been measured. We use general relativistic magnetohydrodynamic simulations, including neutrinos, with several representative viable equations of state, to quantify---for the first time---the effects of the neutron star magnetic field, neutrino emission, and the ejected mass on the linear and nonlinear displacement memory in binary neutron star mergers. We find that the additional contributions due to the emission of electromagnetic radiation, neutrinos and baryonic ejecta can be $\sim 15\%$ of the total memory for moderate magnetic fields and up to $\sim 50\%$ for extreme magnetic fields. The memory is most affected by changes in the equation of state, the binary mass, and the magnetic field. In particular, for moderate premerger field strengths, the dominant impact of the electromagnetic field is the change in the gravitational wave luminosity, and the associated gravitational wave null memory, due to the unstable growth of the magnetic field and the resulting redistribution of angular momentum it induces in the remnant. While the direct electromagnetic contribution to the null memory is additive, the change in the gravitational wave null memory can---in some cases---result in the total memory being \textit{smaller} than that from the corresponding nonmagnetized binary.
Furthermore, in contrast to binary black hole mergers, the growth of the memory in binary neutron star mergers is extended due to the long emission timescale of electromagnetic fields, neutrinos, and ejecta. These results necessitate the consideration of the magnetic field, as well as the equation of state, for accurate parameter estimation in future analyses of gravitational wave memory data.
\end{abstract}

\maketitle

\begin{figure}
    \centering
    \includegraphics[width=\linewidth]{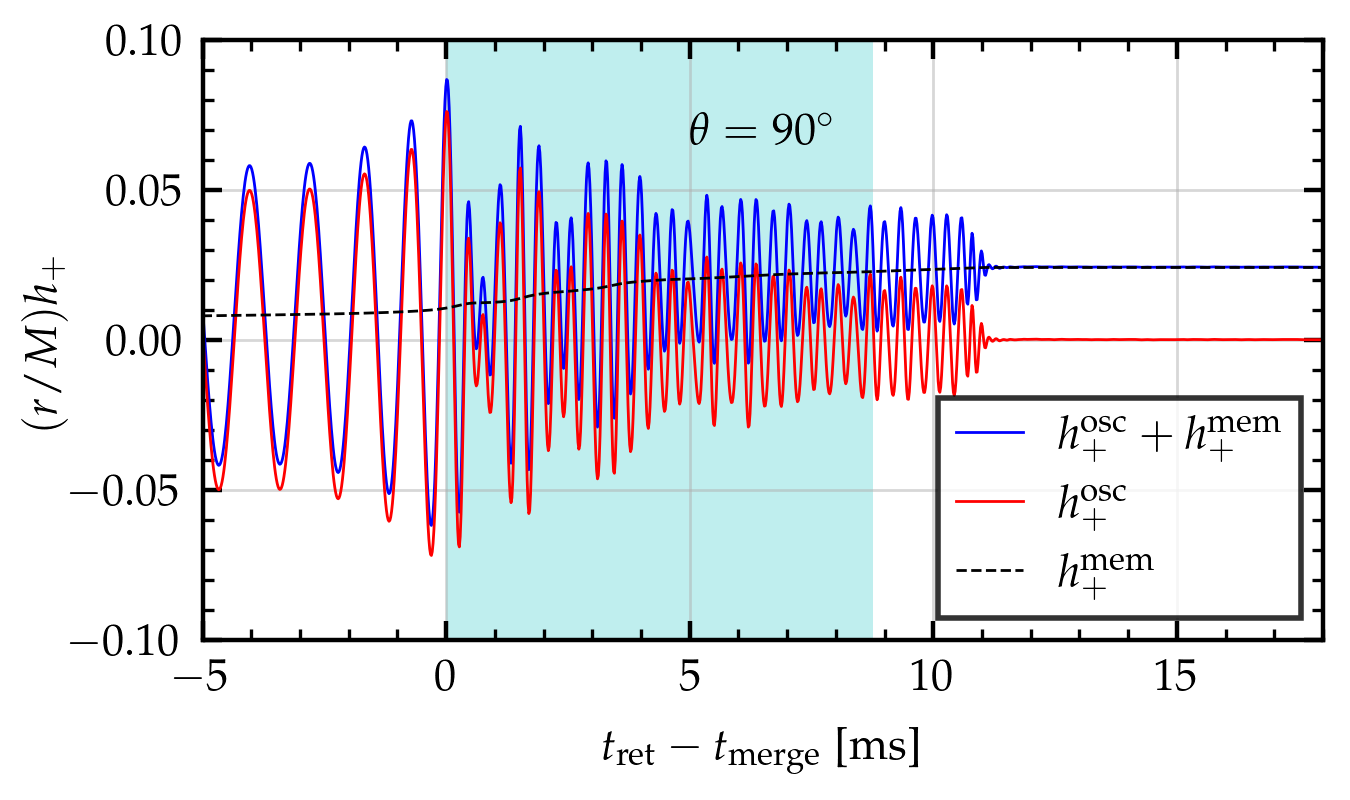}
    \caption{The $h_+$ strain polarization observed in the equatorial plane ($\theta = 90^{\circ}$) showing the oscillatory wave (red), the total displacement memory  (black dashed), and the combined signal (blue) for a BNS simulation with total gravitational mass $M=2.70M_\odot$, mass ratio $q=1$, ALF2 equation of state, and a pulsar-like magnetic field with $\vert B \vert_{\mathrm{max,ins}} = 1.4 \times 10^{16}$G (see main text for details). The memory includes the GW, EM, and ejecta contributions, with the blue rectangle showing the time over which 90\% of the postmerger memory is accumulated.}   
    \label{fig:h_mem_ALF2_2.70_B2.0E15}
\end{figure}

\textit{Introduction.}\textemdash
Despite the tremendous progress in the detection of gravitational waves (GWs), there are still important aspects of general relativity that have not yet been observed, 
and which only recently have been understood. One such phenomenon is the so-called ``memory'' effect, which produces a nonoscillatory contribution to the GW signal (see Fig.~\ref{fig:h_mem_ALF2_2.70_B2.0E15}).  
The first calculation of GW memory was performed by Zel’dovich and Polnarev \cite{Zeldovich1974} who employed the 
linearized Einstein field equations. The memory obtained is thus denoted as the ``linear" memory. 
It results from an overall change of the second time derivative of the matter source's quadrupole moment.
Later, Christodoulou \cite{Christodoulou:1991cr} (also Payne \cite{Payne:1983rrr}
and Blanchet and Damour \cite{Blanchet:1992br}) showed that the nonlinearity of Einstein's equations 
results in an additional memory effect sourced from the cumulative contribution of the effective stress-energy of the GWs themselves. As this arises only at nonlinear order, it was denoted as the ``nonlinear" or ``Christodoulou" memory. Thorne \cite{Thorne:1992sdb} argued that the nonlinear memory can in fact be obtained using the expression for the linear memory, if one replaces the massive particles in the matter source by null gravitons. Wiseman and Will \cite{Wiseman:1991ss} provided a heuristic derivation and a first estimate of the Christodoulou memory for the inspiral phase of coalescing binary systems. Via a semi-analytic approach, Favata \cite{Favata:2009ii} first incorporated the effects of the merger and ringdown. Both also estimated that this effect is as large as $20\mbox{--}30\%$ of the maximum strain (see Fig.~\ref{fig:h_mem_ALF2_2.70_B2.0E15}). In recent years GW memory has been explored in many different ways from a theoretical point of view
\cite{Favata:2008yd,Favata:2009ii,Favata:2010zu,Bieri:2010tq,Bieri:2011zb,Bieri:2013gwa,Bieri:2013ada,Strominger:2014pwa,Winicour:2014ska,Pasterski:2015tva,Nichols2018}. Bieri \etal  ~\cite{Bieri:2010tq,Bieri:2011zb} and Bieri and Garfinkle \cite{Bieri:2013gwa} obtained for the first time the memory  from EM radiation and neutrinos (treated as massless) at nonlinear order, arguing it was equivalent to the Christodoulou memory from GW self-interaction. This motivated a new division of GW memory into ``ordinary" memory arising from non-null matter (e.g. hyperbolic encounters and ejection of baryonic matter) and ``null" memory arising from null radiation (including both EM, neutrinos, and GWs). Here we follow the latter naming convention, terming the null memory from GW, EM and neutrino radiation as ``GW null memory", ``EM null memory" and ``neutrino null memory", respectively. The GW null memory is equivalent to the nonlinear memory discovered by Christodoulou.

Many studies have been performed to numerically compute the memory effect
\cite{Pollney:2010hs,Mitman2020,Mitman:2020bjf,Moxon:2020gha,Moxon:2021gbv,Yoo2023,Chen:2024ieh}. Most of these focused on binary black holes (BBHs) that are expected to have the highest GW luminosities \cite{Lasky2016,Yang2018a,Boersma:2020gxx,Grant2023}. On the other hand, binary neutron stars (BNSs) will have smaller GW luminosities, but they will exhibit additional memory
contributions due to the EM and neutrino luminosities, as well as the baryonic ejecta, all of which are absent in BBH mergers. This includes contributions arising from relativistic jets and (if produced) short $\gamma$-ray bursts (sGRB) \cite{Sago2004,Birnholtz:2013bea,Sakai:2025lks}. In \cite{Tiwari2021,Lopez:2023aja} it is argued that the memory can be used to
distinguish BNS from BBH mergers, while in \cite{Yang:2018ceq} it is argued that it can be used to constrain the yet unknown neutron star (NS) equation of state (EOS). While \cite{Karlson:2018,Yang:2018ceq,Tiwari2021,Lopez:2023aja} investigated the memory from BNS mergers, they only computed the GW null memory. The purpose of this work is to complete the picture by quantifying---for the first time---the additional contributions to the displacement memory from the EM field, neutrinos, and ejecta using GRMHD simulations.

\textit{Simulations and Methods.}\textemdash
Two sets of simulations are used in this work. The first uses our well-tested \textsc{IllinoisGRMHD} code, includes magnetic fields, but has no neutrinos, and is described in detail in \cite{Bamber:2024wqr,Tsokaros:2024wgb}. The second set uses the \textsc{IllinoisGRMHD} thorn \cite{Etienne:2015cea} in the \textsc{Einstein Toolkit} \cite{Loffler:2011ay}, and includes both magnetic fields and neutrino radiation computed via a leakage scheme.

To leading order, the ordinary GW memory in the transverse-traceless (TT) gauge for a system of $N$ bodies with masses $M_A$ and velocities $\boldsymbol{v}_A$, is given by \cite{Braginsky:1987kwo} 
\begin{equation}
    \Delta h^{\rm ord}_{ij} = \frac{4}{r} \;\Delta \sum^N_{A=1} E_A \left[\frac{v^i_A v^j_A}{(1 - \boldsymbol{v}_A \cdot \boldsymbol{N})}\right]^{\mathrm{TT}}, \label{eq:non_rel}
\end{equation}
where $\Delta$ means the difference between the initial and final states of the summation, $E_A = M_A \gamma_A = M_A/\sqrt{1 - v^2_A}$ is the energy of each particle, $\boldsymbol{N}$ is a unit vector that points from the source to the observer, $r$ is the coordinate distance to the source and ``${}^{\mathrm{TT}}$" denotes the transverse-traceless part. Units with $G=c=1$ are used in this work unless stated explicitly.
We use this formula to compute the ordinary memory due to the ejected baryonic mass which we term the ``ejecta memory" (see Supplemental Material). We note here that the ordinary memory is a subset of the linear memory and it was the memory calculated originally by Zel’dovich and Polnarev \cite{Zeldovich1974}.

On the other hand, the memory due to null radiation, the ``null memory", can be written as \cite{Wiseman:1991ss,Thorne:1992sdb}
\begin{equation}
  h^{\rm null}_{ij} = \frac{4}{r} \int^{t_r}_{-\infty} \left[\int\pdv{E}{t'}{\Omega'}\frac{n_i'n_j'}{(1 - \boldsymbol{n}'\cdot\boldsymbol{N})}\dd \Omega' \right]^{\mathrm{TT}}\dd t' , \label{eq:null}
\end{equation}
where $\boldsymbol{n}'$ is a unit radial vector, $\pdv{E}{t'}{\Omega'}$ is the null energy flux per unit solid angle, and $t_r$ is the retarded time. Note that Eq. \eqref{eq:null} can be obtained from Eq. \eqref{eq:non_rel} if you treat the null radiation as made up of unbound particles with velocities $v_A \rightarrow c$ escaping to infinity \cite{Wiseman:1991ss,Thorne:1992sdb}. We use this formula to compute the memory from the GW, EM, and (massless) neutrino null radiation \cite{Epstein1978,Bieri:2013gwa} by substituting their corresponding energy fluxes (see Supplemental Material). Our simulations start 3 to 4 orbits before merger \cite{Bamber:2024wqr}, so we use the matching technique of \cite{Karlson:2018} and the 3PN expressions for the memory from quasicircular inspirals given in \cite{Favata:2008yd} to add the missing inspiral contribution from $t = -\infty$ to the start of the simulation. The simulations end $25$ ms after merger, covering the most interesting time period for BNS mergers.

\textit{Results I: Simulations without neutrinos.}\textemdash
The presence of an EM field has three separate effects on the memory. First, there is the direct contribution: the EM null memory \cite{Bieri:2011zb,Bieri:2013hqa}. Second, there is the effect of the EM field on the postmerger evolution of the remnant (remnants from magnetized BNSs have different properties than those from nonmagnetized progenitors), and hence the GW luminosity and associated memory. Third, there is the effect of the EM fields on the amount of baryonic ejecta and hence the ejecta memory. 

\begin{figure}
    \centering
    \includegraphics[width=\linewidth]{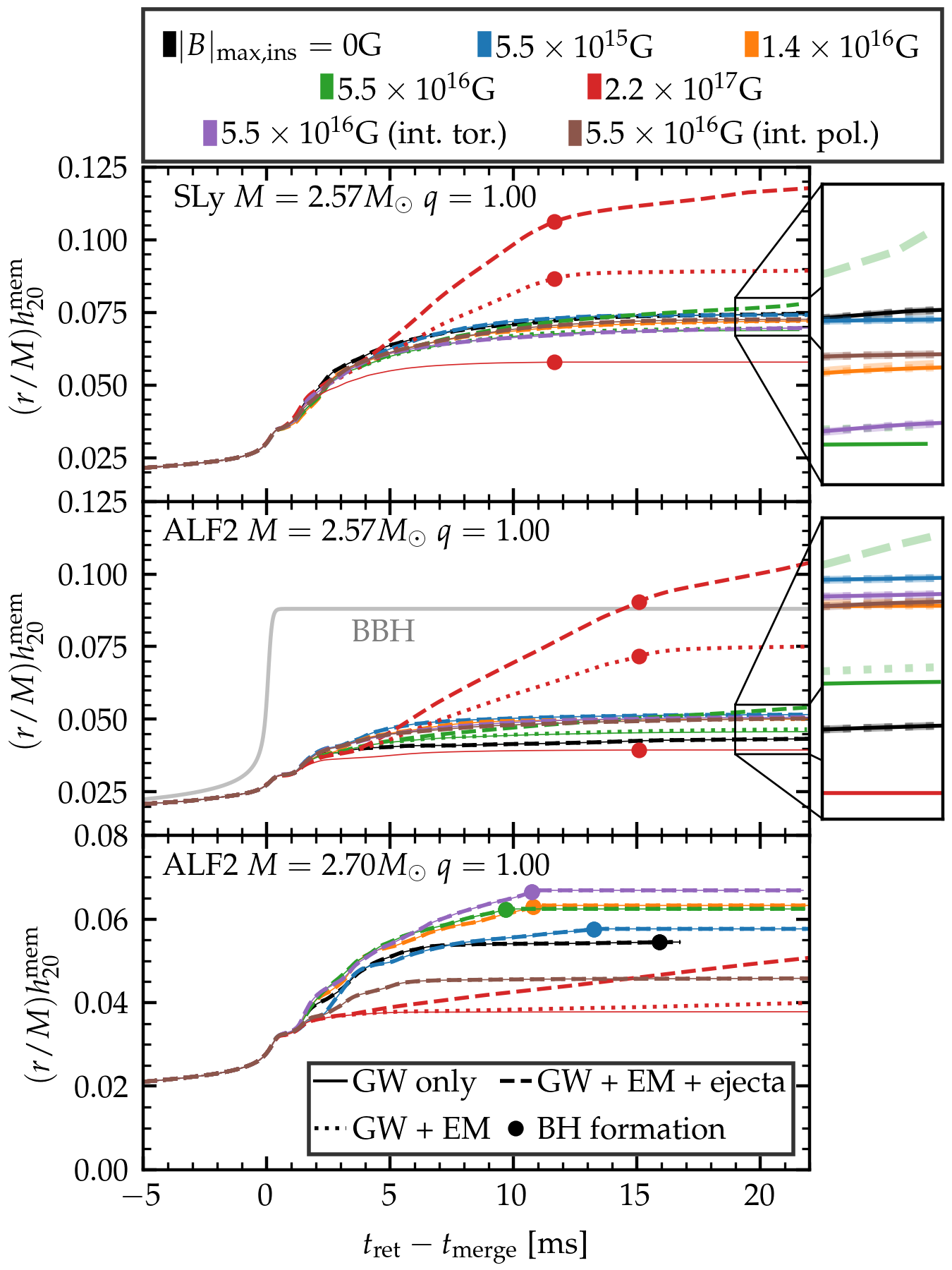}
    \caption{The dominant $l,m=2,0$ mode of the GW memory signal for the first set of simulations conducted with no neutrinos and with mass ratio $q=1$. For comparison, in the central panel we also show the memory from an equal mass BBH with the same ADM mass (gray line). The top two panels also show zoomed-in views of the last few ms for the smaller magnetic field lines.
    }
    \label{fig:h_mem_20_mode}
\end{figure}

In Fig.~\ref{fig:h_mem_20_mode} we show the effect of the EM field on the dominant $l=2,m=0$ mode of the memory (GW, EM, and ejecta) from the first set of simulations (which do not include neutrinos). As in \cite{Bamber:2024wqr,Tsokaros:2024wgb}, the different colors correspond to different inserted magnetic field strengths and topologies, with $\vert B \vert_{\mathrm{max,ins}}$ denoting the maximum field strength at the time of insertion; ``int.~tor." and ``int.~pol." denote initial interior-only toroidal and interior-only poloidal magnetic field topologies (respectively). Inside the stars the initial $B^2/8\pi$ is still $\ll P_c$, the central pressure, so the seed magnetic field is only a small perturbation and is dynamically unimportant. The other cases use a pulsar-like topology \cite{Bamber:2024wqr}. We show the GW null memory with a solid line, the combined GW + EM null memory with a dotted line, and the total memory---including the GW, EM, and ejecta memory contributions---with a dashed line. On each line BH formation is denoted with a solid circle.

The main conclusions from these plots are: 
\begin{enumerate}[topsep=0pt,itemsep=-1ex,partopsep=1ex,parsep=1ex]
      \item The memory from the BNS mergers (at least for these EOSs) is clearly distinct from that of a BBH merger of the same binary mass (grey line, middle panel), due to the slower growth and (for moderate magnetic fields) smaller final amplitude. 
     \item For all but the largest inserted magnetic fields (i.e. $\vert B \vert_{\mathrm{max,ins}}<10^{17}$G), the dominant contribution to the memory is the GW null memory, with the EM and ejecta memory rising to up to $\sim 1\%$ and $\sim 10\%$, respectively, of the GW null memory by the end of our simulations.    
    \item A nonzero magnetic field can have the effect of either \textit{increasing} or \textit{decreasing} the GW null memory. The outcome depends on the strength of the magnetic field, its topology, the NS EOS, and the mass of binary.
\end{enumerate}
Note that BH formation leads to a change of slope of the memory strain amplitude (typically from a larger slope to a flat one, except in exceptionally large magnetic fields; see colored dots in Fig.~\ref{fig:h_mem_20_mode}). It also leads to a reduction in the ejected mass, and hence the ejecta memory, relative to cases with long lived NS remnants. 
    
With regards to point (3), the top and middle panels in Fig.~\ref{fig:h_mem_20_mode} show two binaries with the same mass but different EOS whose memory content behaves differently with respect to the corresponding nonmagnetized runs. In the top panel, all EM fields but the strongest one result in very small differences in the memory relative to the nonmagnetized run. In particular, there is a small monotonic \textit{decrease} in the GW null memory strain for the pulsar-like topology as the magnetic field is increased. 
However, for the same mass with a different EOS (ALF2, middle panel) the GW null memory of the case with smallest non-zero magnetic field $\vert B\vert_{\mathrm{max,ins}} = 5.5\times 10^{15}$G (blue) is \textit{larger} than that of the case with no magnetic field (black). Increasing further the magnetic field leads to a monotonic \textit{decrease} of the GW null memory.
This suggests that even moderate magnetic fields affect the GW null memory of BNS with identical masses differently, depending on the EOS. We also note here that the expectation that magnetized binaries will produce larger total memory signals than nonmagnetized ones due to the additional EM null memory \cite{Bieri:2011zb}, is not necessarily realized, since on one hand the EM null memory is negligible (except for the highest magnetic field case considered here (red)), and on the other, because the postmerger magnetized evolution of the remnant NS can result in a smaller GW null memory. 



The middle and bottom panels of Fig.~\ref{fig:h_mem_20_mode} show results from binaries with the same ALF2 EOS but different masses. In the middle panel, the case with $\vert B\vert_{\mathrm{max,ins}} = 5.5\times 10^{15}$G (blue) has the largest GW null memory, while in the bottom panel it is the case with $\vert B\vert_{\mathrm{max,ins}} = 5.5\times 10^{16}$G and an interior-only toroidal topology (purple). No monotonic behavior with respect to the magnitude of the magnetic field is observed here. Figure \ref{fig:h_mem_20_mode} is consistent with the GW luminosity plots (see Fig.~21 in \cite{Bamber:2024wqr} as well as Fig. \ref{fig:hdot_vs_B}).

The only cases which show significant EM null memory and ejecta memory on the timescale of our simulations are the cases with the strongest inserted magnetic fields, $\vert B \vert_{\mathrm{max},\mathrm{ins}} = 5.5 \times 10^{16}$G and $2.2 \times 10^{17}$G, and a pulsar-like magnetic field topology. These cases have the strongest magnetic fields in the exterior of the NS remnant, which drives both the ejecta outflow and the EM Poynting flux along the axis of the incipient jet (see \cite{Bamber:2024wqr} for more details). Note that in our simulations most of the energy outflow of the incipient jet is in the EM flux, as within our simulation box the EM field has not had time to accelerate the matter to the ultrarelativistic speeds associated with $\gamma-$ray bursts \cite{Bamber:2024kfb}. The cases with $\vert B \vert_{\mathrm{max},\mathrm{ins}} = 5.5 \times 10^{16}$G and interior-only initial magnetic field topology show negligible memory from EM radiation or ejecta compared to the GW null memory, at least on the timescale of the simulation.

\emph{Results II: Simulations with neutrinos.---}Similar to the EM field, neutrinos contribute to the memory in three different ways: directly via the neutrino null memory \cite{Bieri:2013gwa}, by altering the postmerger evolution of the remnant and thus changing the GW null memory, and by changing the ejecta and thus the ejecta memory. However, the effect of the neutrinos on the postmerger evolution is not as prominent as the effect of the magnetic field, at least during the first $\sim 25$ms. 

\begin{figure}
    \centering
    \includegraphics[width=\linewidth]{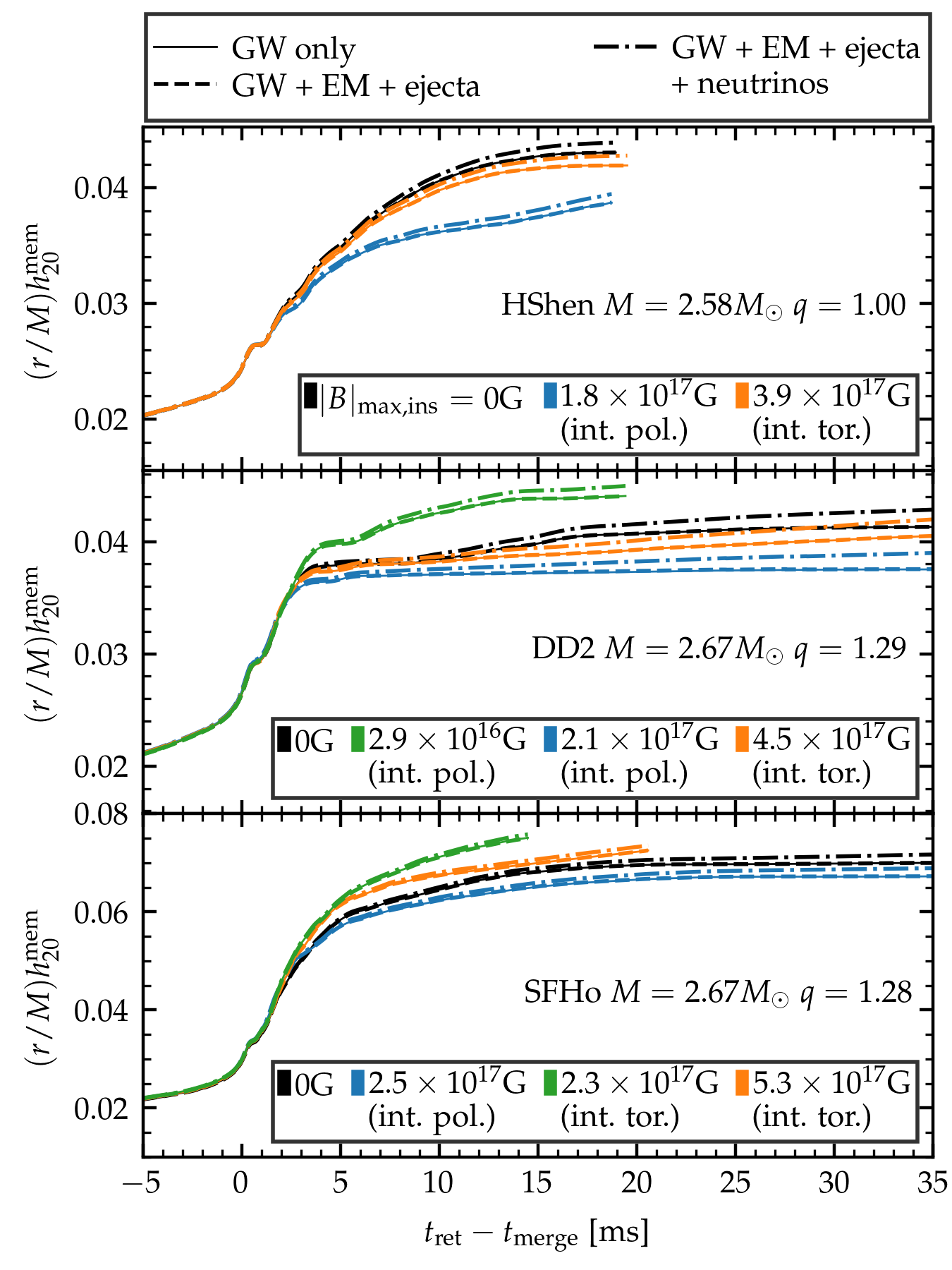}
    \caption{The dominant $l,m=2,0$ mode of the GW memory signal for the second set of simulations, which include interior-only EM fields and neutrino radiation. Note that all of these cases produce supramassive NSs \cite{Cook92b}. Curves are labeled as in Fig. \ref{fig:h_mem_20_mode}.}   
    \label{fig:h_mem_20_mode_ET}
\end{figure}

In Fig.~\ref{fig:h_mem_20_mode_ET} we show the dominant $l=2,m=0$ mode of the memory (GW, EM, ejecta, and neutrino) from the second set of simulations which include interior only EM fields and neutrino radiation. As in the majority of the first set of simulations, the memory is dominated by the GW null contribution, which depends strongly on the EOS (see bottom two panels) and which can either be \textit{larger} or \textit{smaller} than the corresponding nonmagnetized case, depending on the inserted field strength and topology. Although the seed magnetic fields are large, the EM null memory and ejecta memory are negligible, likely because the magnetic field is confined to the interior of the NS. 

In all the cases the neutrino luminosity, with and without magnetic field, rises to $\sim 10^{53}\mathrm{erg}\;\mathrm{s}^{-1}$ roughly $\sim 5\mbox{--}10\,\rm ms$ after merger and persists at roughly that value to the end of the simulations. The emission is anisotropic, as the neutrino flux toward the equatorial region is suppressed: the geometrically thick torus of bound matter is opaque to neutrinos \cite{Dessart:2008zd,Perego:2014fma,Sumiyoshi:2020bdh,Foucart:2024npn}, so it produces a noticeable---although subdominant---memory (on the timescale of the simulations), as one can see from Fig.~\ref{fig:h_mem_20_mode_ET}. As in the first set of simulations, the largest effect arising from the EM field and the neutrinos is the change in the GW null memory due to the redistribution of angular momentum in the remnant induced by the magnetic field (see \cite{Tsokaros:2024wgb} Fig. 1 and associated discussion).

Unlike BBH mergers, the memory in BNS mergers is \textit{extended}: it keeps building up over long timescales postmerger due to the high frequency GW radiation from the postmerger NS remnant \cite{Lopez:2023aja,Bamber:2024wqr} and continued EM, baryonic ejecta, and neutrino emission. However, as both the late-time EM and neutrino memory contributions build up over timescales of seconds or larger \cite{Shapiro:1983,Thompson:2004wi,Lander:2018und,Sekiguchi:2011zd,Hayashi:2024jwt}, they will contribute at sub-Hz frequencies and so will be difficult to detect from ground-based observatories.

\begin{figure}
    \centering
    \includegraphics[width=\linewidth]{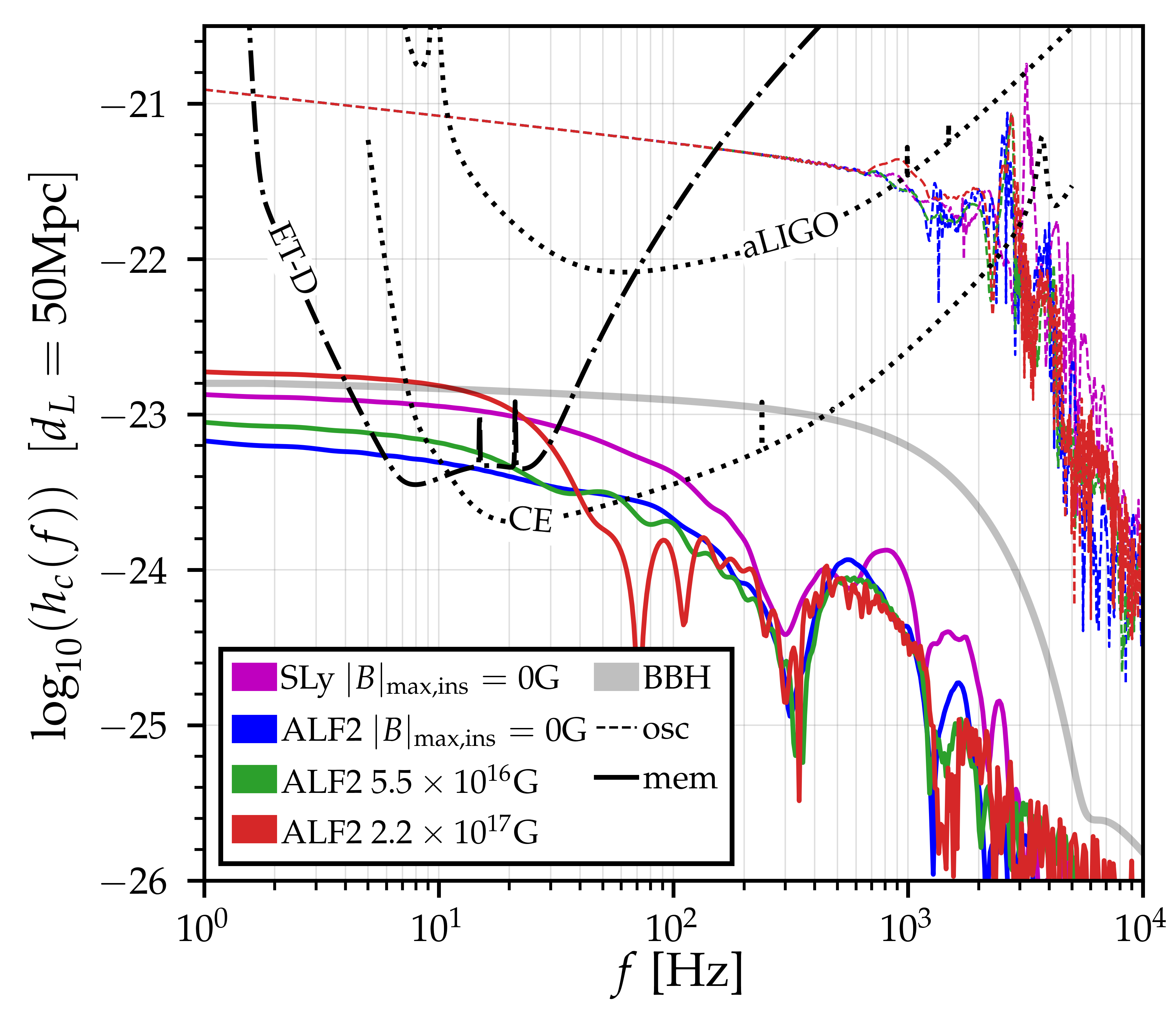}
    \caption{Characteristic strain curves for the oscillatory (dashed) and total memory (solid) parts of the signal for selected binaries with $M=2.57M_{\odot}$ and $q=1$ at a luminosity distance of $50$ Mpc (c.f., top two panels in Fig.~\ref{fig:h_mem_20_mode}). We show zero magnetic field cases for the ALF2 and SLy EOS, and two nonzero (initially pulsar-like) magnetic field cases for ALF2.} We also show the memory for an equal-mass BBH merger with the same mass and distance. Sensitivity curves correspond to Advanced LIGO \cite{aLIGO:2020wna}, the Einstein Telescope (D configuration) \cite{Hild:2010id}, and Cosmic Explorer \cite{Reitze:2019iox,CEcurve}.
    \label{fig:fft_plot}
\end{figure}

\textit{Results III: Detectability.---}Figure~\ref{fig:fft_plot} shows the frequency-domain characteristic strain for both the oscillatory and memory components (including the effect of extended EM emission) for selected $M=2.57M_{\odot}$, $q=1$ simulations. These include three with the ALF2 EOS, one with SLy, and one BBH merger. The memory strain for the BNS mergers (solid colored lines) exhibits a typical form (as seen in, e.g., Fig.~4 of \cite{Johnson:2018xly}), tending to $h_c(f) \sim {\rm const}.$ at low frequencies, a local minimum at $f \sim 100\mbox{--}300$Hz, then a steep drop off at high frequencies. The BNS memory shows less power at high frequencies compared to the BBH, as the BNS memory rises more slowly (c.f., middle panel of Fig.~\ref{fig:h_mem_20_mode} middle).   

While the memory signal is roughly two orders of magnitude smaller than the oscillatory component at the same frequency, it may still be detectable with next-generation GW observatories like the Einstein Telescope and Cosmic Explorer, with a SNR of up to $\sim 5$ for sources at 50 Mpc. Following \cite{Lopez:2023aja}, we can test whether the memory signal alone (i.e. excluding the oscillatory part) can be used to constrain the NS EOS. The mismatch between the memory signals using Cosmic Explorer for the two cases with the same mass $M=2.57M_{\odot}$, no magnetic fields, and two different EOSs (SLy and ALF2) is $0.8\%$, similar to the mismatches observed between different EOSs in Table I of \cite{Lopez:2023aja}. Applying the waveform distinguishability criterion (e.g., Eq.~(44) of \cite{Bamber:2024wqr}) suggests that $\sim 30\,{\rm Mpc}$ is the maximum distance at which those two cases can be distinguished.

The most extreme (and somewhat unlikely \cite{Bamber:2024wqr}) magnetic field strength we explored ($\vert B \vert_{\mathrm{max,ins}}=2.2\times 10^{17}$G, red) shows a significantly larger strain at low frequencies compared to the $B=0$ case with the same EOS and mass (dark blue). This is due to the extra memory from the EM emission and ejecta. Note that a BNS---magnetized or not---can in principle be distinguished from a BBH with the same mass by the existence of one or more local minima in the characteristic strain (a BBH has no local minimum). Besides the local minimum at $\sim 300\,{\rm Hz}$, an additional local minimum appears at $\sim 50\mbox{--}100$Hz in the highest field strength case. This second minima could \textit{in principle} be used to indicate the presence of strong magnetic fields. In practice, however, this will be difficult, since the local minima fall below the sensitivity curves of next generation GW observatories. The cases with more realistic magnetic field strengths (such as the pulsar-like $\vert B \vert_{\mathrm{max,ins}}=5.5\times 10^{16}$G case, green) show smaller differences. However, the mismatches using Cosmic Explorer between the $B=0$ and $\vert B \vert_{\mathrm{max,ins}}=5.5\times 10^{16}$G cases with the same mass and EOS are $0.3\%$ and $1.2\%$ for SLy and ALF2 respectively; this is comparable to the mismatch of $0.8\%$ between the SLy and ALF2 ($B=0$) cases. Hence, if future GW observatories do obtain sufficient sensitivity to make EOS inferences from the memory waveform, these results suggest that strong magnetic fields may be degenerate with EOS effects. This is analogous to the case of the postmerger frequency spectrum from the oscillatory waves \cite{Tsokaros:2024wgb}.

\textit{Conclusions.}\textemdash
We performed the first memory computations for BNS mergers that include not only the GW null memory, but the EM, neutrino, and ejected matter memory contributions as well. EM fields can either \text{increase} or \text{decrease} the memory, depending on the field strength, its topology, EOS, and binary mass. But the EM null memory is subdominant to the GW null memory for all but the largest EM fields. The memory from the ejected matter and neutrinos is also subdominant, but can be nonnegligble for certain EOS, masses, and field strengths. Future analyses of BNS memory should also consider degeneracies between different physical phenomena, including the magnetic field and the EOS \cite{Tsokaros:2024wgb}. Movies and additional visualizations highlighting our results can be found at \cite{website}.

\textit{Acknowledgments.}\textemdash
We thank Lydia Bieri and Alan Wiseman for useful discussions.
This work was supported in part by National Science Foundation (NSF) Grants No.
PHY-2308242, No. OAC-2310548 and No. PHY-2006066 to the University of Illinois
at Urbana-Champaign; and PHY-1653374 to Montclair State University. A.T. acknowledges support from the National Center for
Supercomputing Applications (NCSA) at the University of Illinois at
Urbana-Champaign through the NCSA Fellows program.  M.R. acknowledges support
by the Generalitat Valenciana Grant CIDEGENT/2021/046 and by the Spanish
Agencia Estatal de Investigaci\'on (Grant PID2021-125485NB-C21).  
Further support has been provided by the EU's Horizon 2020 Research and Innovation
(RISE) programme H2020-MSCA-RISE-2017 (FunFiCO-777740) and by the EU Staff Exchange 
(SE) programme HORIZON-MSCA-2021-SE-01 (NewFunFiCO-101086251).
This work used Stampede2 at TACC and Anvil at Purdue University through allocation
MCA99S008, from the Advanced Cyberinfrastructure Coordination Ecosystem:
Services \& Support (ACCESS) program, which is supported by National Science
Foundation grants \#2138259, \#2138286, \#2138307, \#2137603, and \#2138296.
This research also used Frontera at TACC through allocation AST20025. Frontera
is made possible by NSF award OAC-1818253.  The authors thankfully acknowledge
the computer resources at MareNostrum and the technical support provided by the
Barcelona Supercomputing Center (AECT-2023-1-0006).


\bibliography{combined_bib}

\section{Supplemental Material}

\textit{Simulations and Methods.}\textemdash
\label{sup}
The second set of GRMHD simulations consist of NSNS mergers of irrotational NSs with three different hot tabulated EOSs (HShen \cite{Shen:2011qu}, DD2 \cite{Hempel:2009mc} and SFHo \cite{Steiner:2012rk}). The HShen cases are equal mass with ADM mass $M = 2.58 M_{\odot}$, while the DD2 and SFHo cases are unequal mass: $q=1.29, M=2.67 M_{\odot}$ and $q = 1.28, M=2.67 M_{\odot}$, respectively. All produce supramassive NS remnants that persist to the end of the simulations. For each group we explored two magnetic field topologies: i) an interior-only toroidal magnetic field with an initial maximum value $\sim 4\mbox{--}5  \times 10^{17}\,\rm{G}$ (having the maximum plasma parameter $\beta^{-1}:= p_{\rm mag}/p_{\rm gas}= 0.003125$ as in previous simulations~\cite{Ruiz:2020via,Sun:2018gcl}); and ii) an interior-only poloidal magnetic field with an initial maximum value $\sim 2 \times 10^{17}\,\rm{G}$. The initial $B^2/8\pi$ is still $\ll P_c$, the central pressure, so the seed magnetic field is only a small perturbation and is dynamically unimportant. For the DD2 (SFHo) cases we conducted an additional simulation with a poloidal (toroidal) topology and a weaker initial magnetic field of $\sim 3 \times 10^{16}\,\rm{G}$. The spatial resolutions used are $\Delta x_{\mathrm{min}} = 277\,\rm{m}, 221\,\rm{m}, 245\,\rm{m}$ for the HShen, DD2 and SFHo cases respectively. 
The evolution of the GRMHD equations coupled to neutrino radiation is done via a leakage scheme that incorporates neutrino production via $\beta$-processes, $e^{-}$--$e^+$ pair annihilation, transverse plasmon decay, and nucleon-nucleon bremsstrahlung (for further details see \cite{Werneck:2022exo} Secs.~II and III).

We extract the spin-weighted spherical harmonic modes of the Weyl scalar $\Psi_4$ over spherical surfaces at a radius $r \sim 280M$, then use the relation $\Psi_4 = \ddot{h}_+ - i\ddot{h}_{\times}$ (valid in the wave zone) to convert to $h_{+,\times}$ strain polarizations. To minimize the contamination from random noise, the double time integral is done in the Fourier domain with a high-pass filter \cite{Bishop:2016lgv,Reisswig:2010di}; this means we only obtain the oscillatory strain component. To reconstruct the nonoscillatory memory due to the GW emission we use the methods of \cite{Favata:2009ii,Favata:2010zu,Karlson:2018}. The strain is decomposed into harmonic modes as $h_+ - ih_{\times} = h = \sum_{lm}h_{lm}{}_{-2}Y^{lm}(\theta,\phi)$, where ${}_{s}Y^{lm}(\theta,\phi)$ are spin $s$ weighted spherical harmonics~\cite{Ruiz:2007yx}. From Eq.~\eqref{eq:null} the null memory satisfies
\begin{equation}
    r\dot{h}^{\mathrm{null}}_{lm} = 16\pi 
    \sqrt{\tfrac{(l-2)!}{(l+2)!}} \int \pdv{E}{t'}{\Omega'}(\theta',\phi')\;{}_{0}Y^{lm*}(\theta',\phi')\;\dd \Omega'. \label{eq:null_lm}
\end{equation}
The GW energy flux is given by 
\begin{equation}
\pdv{E_{\mathrm{GW}}}{t}{\Omega} = \frac{r^2}{16\pi}\langle \vert \dot{h} \vert^2\rangle \approx \frac{r^2}{16\pi} \vert \dot{h} \vert^2 \ ,
\end{equation}
so we obtain the GW null memory as     
\begin{equation}
\begin{split}
    r\dot{h}^{\mathrm{GW}}_{lm} =& 
    \;r^2\sqrt{\tfrac{(l-2)!}{(l+2)!}}\sum^{l_{\mathrm{max}}}_{l',l''=2}\ 
                                      \sum^{m'=l'}_{m'=-l'}\ 
                                      \sum^{m''=l''}_{m''=-l''}\dots \\
    &\dots (-1)^{m'+m''}\dot{h}_{l'm'}\dot{h}^*_{l''m''}G^{2\;-2\;0}_{l'\;l''\;l\;m'\;-m''\;-m} \ , \label{eq:hmem_dot}
\end{split}
\end{equation}
where $l_{\mathrm{max}} = 4$ is the maximum $l'$ and $l''$ where we truncate the sums, and 
\begin{equation}
\begin{split}
    &G^{s_1\;s_2\;s_3}_{l_1 l_2 l_3 m_1 m_2 m_3} := \\ &\int {}_{-s_1}Y^{l_1 m_1}(\theta,\phi) \,{}_{-s_2}Y^{l_2 m_2}(\theta,\phi) \,{}_{-s_3}Y^{l_3 m_3}(\theta,\phi)\,\dd \Omega
\end{split}
\end{equation}
[see \cite{Karlson:2018} Eqs.~(2.1.5)--(2.1.7) and \cite{Favata:2008yd} Eqs.~(2.32) \& (3.3)].
As the memory itself contributes a negligible component of the GW energy flux, we can safely approximate the $\dot{h}_{lm}$ terms on the right hand side of \eqref{eq:hmem_dot} with just the oscillatory component. 

\begin{figure}
    \centering
    \includegraphics[width=\linewidth]{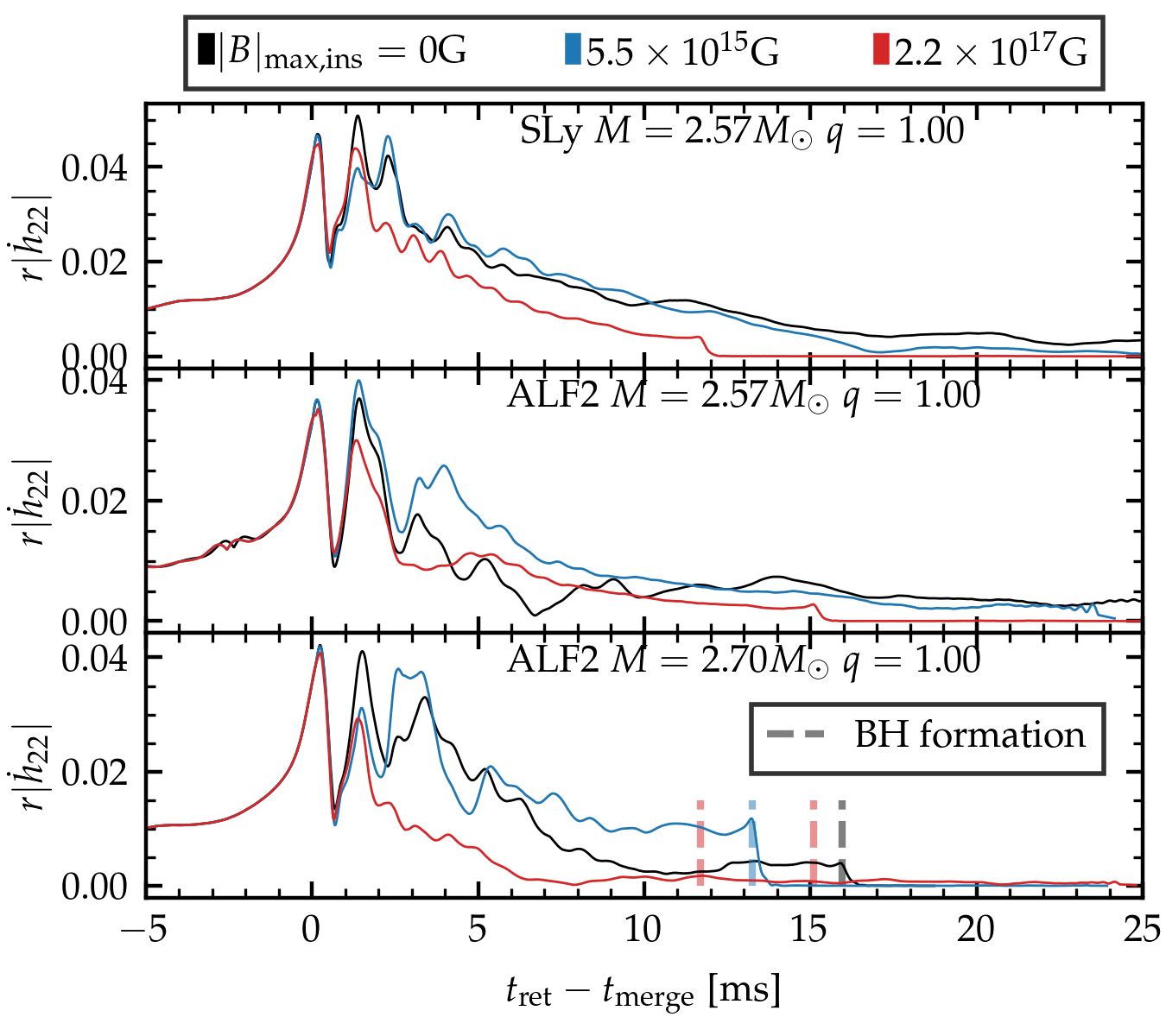}
    \caption{Magnitude of the time derivative of the $l=m=2$ mode of the complex strain for three different initial magnetic field strengths and pulsar-like initial magnetic field topology.}
    \label{fig:hdot_vs_B}
\end{figure}

Our simulations begin at a finite time $t_0$; however, the GW null memory will build up over the entire inspiral from $t = -\infty$. Therefore, we add the contribution from $t < t_0$ using the 3PN expressions for the displacement memory for quasicircular orbits given in \cite{Favata:2009ii} using the matching technique described in \cite{Karlson:2018} Sec.~(2.3). For a binary in the $x\mbox{--}y$ plane, only the $m=0$ and even $l$ modes have memory, and only in the $h_{+}$ polarization. The dominant modes for quasicircular inspirals are $h_{20}$ and $h_{40}$, with negligible contributions from $h_{l0}$ for $l \geq 6$ \cite{Favata:2009ii,Karlson:2018}. 
The dominant contribution to the GW luminosity and hence to the GW null memory in Eq. \eqref{eq:hmem_dot} is from the $\dot{h}_{2\pm 2}$ terms. In Fig.~\ref{fig:hdot_vs_B} we show the magnitude of $r\dot{h}_{22} = r(\dot{h}^{+}_{22}-i\dot{h}^{\times}_{22})$ for a selection of the first set of merger simulations (without neutrinos) with pulsar-like initial magnetic fields of different initial strengths. One can see the substantial difference in the postmerger amplitude between the different cases, which in turn produces the non-monotonic (with respect to the magnetic field) behavior of the GW null memory as shown in Fig. \ref{fig:h_mem_20_mode}. 

The EM null memory is obtained by substituting the EM luminosity into Eq.~\eqref{eq:null_lm}. For the neutrino radiation, we calculate the total luminosity using \cite{Shibata:2007gp}
\begin{equation}
    L_{\nu_i} = -\int  \dot{Q}_{\nu_i} u_t \sqrt{-g} \;\dd^3 x, \label{eq:L_nu}
\end{equation}
where $\nu_i$ is the neutrino species and $\dot{Q}_{\nu_i}$ is the emissivity. We assume the angular distribution of the radiation follows Fig.~5 in \cite{Sumiyoshi:2020bdh} which was obtained by solving the 6D Boltzmann equation for radiation from a neutron star remnant.

We can obtain the ordinary memory from the nonrelativistic ejecta from Eq. \eqref{eq:non_rel}. For a continuum distribution of unbound ejected matter we can write it as
\begin{equation}
    r h^{\mathrm{ej}}_{ij} = 4 \;\int_{r>r_{\mathrm{esc}}} \gamma \rho_{\mathrm{ej}} \left[\frac{v_i v_j}{(1 - \boldsymbol{v} \cdot \boldsymbol{N})}\right]^{\mathrm{TT}} \dd V,
\end{equation}
where $\rho_{\mathrm{ej}} = \rho_0 H(-u_t-1)H(v^r)$, $\rho_0$ is the rest-mass density, $v^r$ is the radial component of the velocity, $r_{\mathrm{esc}}$ is a minimum escape radius and $H$ is the Heaviside function. The transverse-traceless (TT) projection operator is $\Pi^{kl}_{ij} = P^k_i P^l_j - \tfrac{1}{2}P_{ij}P^{kl}$, where $P_{ij} = \delta_{ij} - N_i N_j$ and the plus and cross projection operators are $e^{+}_{ij} = \tfrac{1}{2}(P_i P_j - Q_i Q_j)$ and $e^{\times}_{ij} = \tfrac{1}{2}(P_i Q_j + Q_i P_j)$. Here $P_i = e^{\Theta}_i$, $Q_i = e^{\Phi}_i$, and $(\Theta,\Phi)$ is the angular direction of $\boldsymbol{N}$ in spherical polar coordinates [see \cite{Favata:2008yd}, Eqs.~(2.2)\mbox{--}(2.6)]. We can get $h = h_{+} - ih_{\times}$ from $h = \tfrac{1}{2}R^i R^j h^{\mathrm{TT}}_{ij}$, where $R_i = (P_i - i Q_i)$. Now $\tfrac{1}{2}R^i R^j \Pi^{kl}_{ij} = \tfrac{1}{2}R^k R^l$, so 
\begin{equation}
    r h^{\mathrm{ej}}_{ij} = 4 \;\int \gamma \rho_{\mathrm{ej}} \int \frac{\tfrac{1}{2}R^i R^j v_i v_j}{(1 - \boldsymbol{v} \cdot \boldsymbol{N})} {}_{-2}Y^{*lm}(\Theta,\Phi) \;\dd \Omega \;\dd V, \\
\end{equation}
where $\dd \Omega = \sin\Theta\, \dd \Theta\, \dd \Phi$. For the first set of binary mergers with mass ratio $q=1$, we approximate the ejecta as axisymmetric with $r_{\mathrm{esc}} = 30M$. For the second set, including non-equal mass mergers, we use Lagrangian tracer particles to track the ejecta angular distribution. 

\textit{Estimates}\textemdash
For the nonrelativistic baryonic ejecta the final memory amplitude can be estimated as \cite{Lopez:2023aja,Braginsky:1987kwo}
\begin{equation}
    \frac{rh^{\mathrm{ej}}}{M} \sim \frac{2 M_{\mathrm{ej}} v^2_{\mathrm{ej}}}{M} \sim 10^{-3} \left(\frac{M_{\mathrm{ej}}}{0.1 M_{\odot}}\right)\left(\frac{v_{\mathrm{ej}}}{0.2c}\right)^2,
\end{equation} where $M_{\mathrm{ej}}$ and $v_{\mathrm{ej}}$ are the final mass and typical velocity of the ejecta. For typical astrophysical NSNS mergers this is expected to be much smaller than the GW null contribution. However, in case of very strong ($\vert B \vert_{\mathrm{max,ins}} = 2.2\times 10^{17}\,{\rm G}$) pulsar-like magnetic fields, the large estimated final ejecta mass $M_{\mathrm{ej}} \sim 0.2M_{\odot}$ and large  ejecta velocity $v_{\mathrm{ej}} \sim 0.5c$ \cite{Bamber:2024wqr} give $rh^{\mathrm{ej}}/M$ up to 84\% of the GW null memory. 

The EM Poynting luminosity increases postmerger as the magnetic field is amplified due to magnetohydrodynamic instabilities, dynamo , and magnetic winding in the NS remnant and/or disk and collimates above the poles, forming incipient jets or jet-like structures (see \cite{Bamber:2024kfb} and references therein, and \cite{Kiuchi:2022nin,Kiuchi:2023obe,Hayashi:2024jwt} for long-timescale simulations). 
The rise time and steady-state luminosity depends on the initial inserted magnetic field strength and topology \cite{Bamber:2024wqr}, ranging from a few to 10s of ms and $10^{47}\mbox{--}10^{54}\,\mathrm{erg}\,\mathrm{s}^{-1}$ in our simulations. For BH remnants this emission may persist for seconds \cite{Hayashi:2024jwt}, while long-lived NS remnants have spindown timescales due to dipole radiation of $\tau_{\mathrm{SD}} \sim 10^{3}B^{-2}_{15}\;T^2_{\textup{ms}}\;\textup{s}$ \cite{Shapiro:1983,Thompson:2004wi,Lander:2018und}, where $B_{15}$ is the magnetic field strength in units of $10^{15}\,\rm{G}$ and $T_{\textup{ms}}$ is the rotation period in ms. The final EM null memory strain obtained from Eq.~\eqref{eq:null} for such a long-lived NS remnant is then
\begin{equation}
    \frac{rh^{\mathrm{EM}}}{M} \sim 10^{-2}\left(\frac{L_{\mathrm{EM}}}{10^{50}\,\mathrm{erg}\,\mathrm{s}^{-1}}\right)\left(\frac{\tau_{\mathrm{SD}}}{10^3\,\mathrm{s}}\right).
\end{equation}

The neutrino emission from the hot NS remnant persists with luminosity $L_{\mathrm{\nu}} \sim 10^{53}\mathrm{erg}\,\mathrm{s}^{-1}$ for cooling timescales of $\tau_{\mathrm{cool}} \sim 2\mbox{--}3\,{\rm s}$ \cite{Sekiguchi:2011zd}, while collapse to a BH triggers a power law decrease in the luminosity from the remaining disk \cite{Siegel:2017nub,Kiuchi:2022nin}. For long-lived NS remnants, assuming the emission retains a similar angular dependence, we obtain from Eq.~\eqref{eq:null} a final strain memory of
\begin{equation}
  \frac{rh^{\mathrm{neutrinos}}}{M} \sim 2 \times 10^{-2} \left(\frac{L_{\nu}}{10^{53}\,\mathrm{erg\,s}^{-1}}\right)\left(\frac{\tau_{\mathrm{cool}}}{3\,\mathrm{s}}\right).   
\end{equation}

\end{document}